\documentstyle[mprocl,psfig]{article}

\def\today{25 February 1998}

\newcommand\prbagt{\mathrel{\raise.3ex\hbox{$>$}\mkern-14mu
             \lower0.6ex\hbox{$\sim$}}}
\newcommand\prbalt{\mathrel{\raise.3ex\hbox{$<$}\mkern-14mu
             \lower0.6ex\hbox{$\sim$}}}

\begin{document}

\title{Critical phenomena in gravitational collapse\footnote{To appear
in {\it Proceedings of the 8th Marcel Grossmann Meeting}, edited by
T. Piran (World Scientific, Singapore).}}

\author{Patrick R Brady}
\address{Theoretical Astrophysics 130-33,
California Institute of Technology,\\ Pasadena, CA 91125}

\author{Mike J Cai}
\address{Department of Physics,  University of California,\\
Berkeley,  CA 94720} 
\maketitle
\abstracts{
\centerline{(\today)}\vskip 0.1in
This article reviews classical and quantum aspects of critical
phenomena in gravitational collapse.  We pay special attention to the
origin of the scaling law for black hole mass, and to phase
transitions in which black hole formation turns on at finite mass.  We
present some new results for perfect fluids with pressure proportional
to density.  }

%
%
\section{Introduction}
\label{prb-s10}

The discovery, by Choptuik,\cite{Choptuik_M:1993} of critical point
behavior in gravitational collapse is one of the most significant
achievements of numerical relativity to date.  Efforts to understand
the Choptuik phenomena have opened an exciting area of research in
general relativity, one which has benefited from the interplay between
mathematical and numerical investigations.%
\cite{Koike_T:1995}$^{\!-\,}$\cite{Hod_S:1997}

The problem addressed by Choptuik was suggested by the work of
Christodoulou~\cite{Christodoulou_D:1986,Christodoulou_D:1991} on the
gravitational collapse of spherically symmetric configurations of
massless, minimally coupled scalar fields.  Christodoulou proved the
global existence and uniqueness of regular solutions for initial
scalar field configurations in a neighborhood of trivial initial data
({i.e.,} flat, empty space).  When the initial data is strong, in a
precise sense, he also established that the black hole mass has a
positive lower bound.\cite{Christodoulou_D:1991} Choptuik examined
the intermediate regime numerically.  By constructing interpolating
families of solutions depending on a single parameter $\eta$, Choptuik
demonstrated that there generally exists a critical value $\eta_*$
such that $S[\eta<\eta_*]$ are solutions in which the scalar field
disperses to infinity, and $S[\eta>\eta_*]$ are solutions in which the
field collapses to form a black hole; he called these {\em
subcritical} and {\em supercritical} solutions, respectively.  He also
observed two important properties of near critical evolutions
S[$\eta \approx \eta_*$]:
\begin{enumerate}
\item For each family of initial data, the masses of black holes formed 
in marginally supercritical collapse obey a scaling relation
\begin{equation}
	M_{\rm BH} = K|\eta-\eta_*|^\beta \label{prb-e:mscale}
\end{equation}
where $\beta \approx 0.37$ is universal, {i.e.,} independent of the
initial data,  although the constant $K$ is a family dependent.
\item The evolution of the scalar field is characterized by
scale-invariant echoing with a period, in logarithmic time, given by
$\Delta \approx 3.4$.
\end{enumerate}
These two observations led Choptuik to speculate that precisely
critical evolutions approach a unique solution in which
scale-invariant echoes accumulate at a massless, central singularity.

It is now known that critical point behavior is a general feature of
gravitational collapse.  Consider the evolution of a physical system
consisting of matter coupled to gravity, or simply of the
gravitational field itself.  One can think of the equations which
govern this system as a map $\cal E$ from the space of initial data
into the space of solutions.  Suppose the strength of the initial data
$I[\eta]$ is characterized by the single tunable parameter $\eta$
(taken to be in the range $0\leq \eta \leq \infty$ for concreteness),
then one constructs a family of solutions $S[\eta]$ by
\begin{equation}
 {\cal E}:I[\eta] \rightarrow S[\eta] \; .
\end{equation}
Provided weak cosmic censorship~\footnote{By weak cosmic censorship we
mean that, in the evolution of generic initial conditions,
singularities are always hidden behind an event horizon.} holds,  
critical point behavior may be expected whenever these solutions
interpolate between black hole formation and stable, regular
solutions.\footnote{Minkowski spacetime is a special case where the
material disperses completely}   Some of the questions which then
arise are:
\begin{enumerate}
\item Is cosmic censorship upheld in all families of interpolating solutions?
\item What is the origin of the scaling relation for the black hole mass?
\item Does black hole formation always turn on at infinitesimal mass?
\item Is the critical point always associated with scale invariant
solutions?
\item What are the semi-classical corrections to the critical point
behavior?
\end{enumerate}
In this contribution, we provide only a flavor of the work that has
been done to answer these, and other questions.  Our presentation is
not intended to be complete, rather it focuses on issues that were
discussed in the parallel session at the 8th Marcel Grossmann meeting
(MG8).  A more complete review, including technical details of the
many interesting results which have been obtained, has recently been
prepared by Gundlach.\cite{Gundlach_C:1997d}

Nevertheless, we would be remiss if we did not mention some of the
notable contributions which were not represented at MG8.  All of the
work to date has been restricted to spherical symmetry except for the
study of axisymmetric collapse of pure gravitational waves by Abrahams
and Evans.\cite{Abrahams_A:1994} They demonstrated the existence of
critical point behavior and derived a scaling law for the black-hole
mass similar to that in Eq.~(\ref{prb-e:mscale}) with $\beta\approx
0.37$.  Despite the extreme complexity of the analysis, they also
presented tentative evidence for scale invariant echoing of the metric
on the symmetry axis of the spacetime.  The black-hole mass scaling
exponent is intriguingly similar to that observed in scalar field
collapse, although the echoing period, $\Delta \approx 0.6$, of the
solution is quite different.

The initial explanation of the scaling relation (\ref{prb-e:mscale}) for
black hole mass was made possible by the work of Evans and
Coleman.\cite{Evans_C:1994} They showed that the critical solution, in
gravitational collapse of radiation fluid, is continuously
self-similar by examining one parameter families of interpolating
solutions, {\em and} direct construction of the intermediate
self-similar attractor.  They indicated that the critical exponent
might be derived by considering linearized perturbations about the
critical solution.  This suggestion was followed up by Koike~{\it et
al.}~\cite{Koike_T:1995} who argued that the critical exponent,
determined to be $\beta\approx 0.36$ by Evans and Coleman, is directly
related to the largest Lyapunov exponent of perturbations about the
critical solution. (This idea was also explored by Eardley and
Hirschmann~\cite{Hirschmann_E:1995} in the context of complex scalar
field collapse.)  In this way Koike~{\it et al.} provided the first
direct computation of the critical exponent as $\beta=0.355$ for
radiation fluid.  Maison~\cite{Maison_D:1996} then extended these
results to more general equations of state for the perfect fluid.

It is worth summarizing the argument which relates the Lyapunov
exponent to the critical exponent;  the discussion
follows Maison.\cite{Maison_D:1996}  The general spherically
symmetric line element can be written as
\begin{equation}
	ds^2 = - \alpha^2 dt^2 + f^{-1} dr^2 + r^2 (d\theta^2 +
	\sin^2\theta \, d\phi^2) \; .
	\label{prb-e:line-element}
\end{equation}
The self-similar critical solution has $\alpha = \alpha_*(\xi)$, and
$f=f_*(\xi)$, where $\xi=r/t$ and the time coordinate is normalized so
that $t=0$ at the singularity in the critical solution.  An important
feature of this solution is the existence of a sonic point $\xi_{\rm
sp}$ where the fluid moves at the speed of sound relative to $\xi =
\xi_{\rm sp}$.  Now, consider the linear stability of these solutions
against perturbations with the general form
\begin{equation}
	-(\eta - \eta_*) f_1(\xi) t^{-\lambda} \; ,
\end{equation}
where $\lambda$ is a positive constant.  These perturbations arise
as small deviations from precisely critical initial data ---
consequently, they are proportional to $(\eta-\eta_*)$.  There is a
unique value of $\lambda$ such that the perturbations are regular both
at $r=0$ and at $\xi_{\rm sp}$.  An apparent horizon in the perturbed
spacetime is given by $(\xi_h,t_h)$ such that
\begin{equation}
	f_* (\xi_h) - (\eta-\eta_*) f_1(\xi_h) t_h^{-\lambda} = 0 \; .
	\label{prb-e:app-horizon}
\end{equation}
The mass $M_{\rm BH}$ inside the apparent horizon is related to its
radius $r_h$ by $M_{\rm BH} = r_h/2$. Substituting $t_h = r_h/\xi_h$
into Eq.~(\ref{prb-e:app-horizon}),  we can solve for $M_{\rm BH}$ as
\begin{equation}
	M_{\rm BH} = r_h/2 \propto |\eta-\eta_*|^{1/\lambda} \; .
\end{equation}
The scaling exponent is then read off as $\beta=1/\lambda$.

In addition to these works, Gundlach has made significant
contributions to the analytic understanding of critical phenomena when
the critical solutions have self-similar echoes.  He has termed this
symmetry {\em discrete self-similarity} (DSS), and has constructed
solutions with DSS by writing the metric as a Fourier series, and
recasting Einstein's equations as an eigenvalue
problem for the echoing period.\cite{Gundlach_C:1995}  As in the case
of perfect fluids, where Maison was able to predict the critical
exponent for situations which had not been numerically explored, the
power of Gundlach's analysis was demonstrated by his {\em
prediction}~\cite{Gundlach_C:1997} that there should be periodic fine
structure on the mass scaling relations such that
\begin{equation}
M_{\rm BH} \propto |\eta-\eta_*|^\beta e^{\Psi( \ln |\eta-\eta_*| )} \; ,
\end{equation}
where $\Psi$ has period $\Delta/(2\beta) \simeq 4.61$ in $\ln|\eta-\eta_*|$.
Similar arguments were presented independently by Hod and
Piran~\cite{Hod_S:1997a} who confirmed the existence of this effect
numerically.  

In collaboration with Martin-Garcia, Gundlach considered charged
scalar field collapse and again correctly predicted the scaling
relation that has been observed for the final black hole
charge.\cite{Gundlach_C:1996a} Recently he has extended his analysis
to include small deviations from spherical symmetry, and has derived a
scaling relation for the angular momentum parameter of spinning black
holes near the critical point.\cite{Gundlach_C:1997b,Gundlach_C:1997c}

The remainder of this paper is organized as follows.  In
subsection~\ref{prb-ss:massless}, we present a simple argument for the
periodic fine structure discovered by Gundlach,\cite{Gundlach_C:1997}
and independently by Hod and Piran.\cite{Hod_S:1997} Interesting new
phenomenology is observed when scale invariance of the underlying
mathematical equations is broken.  This was first demonstrated in the
work of Choptuik~{\it et al.} by considering the gravitational
collapse of a Yang-Mills field where the SU(2) charge introduces a
fundamental scale into the problem.\cite{Choptuik_M:1996} In
subsection~\ref{prb-ss:massive} we discuss similar results for massive
scalar field collapse; in particular, we focus on a simple criterion to
determine when the mass term is important in critical evolutions.  We
then outline the results which have been obtained by Choptuik~{\it et
al.} who have shown that the {\it approximate black hole} solution
discovered by Van~Putten~\cite{Vanputten_M:1996a} is unstable.  This
result also has implications for massless scalar field collapse since
van Putten's solution may lie at the threshold of black hole
formation.  Some preliminary results from a study of fluid collapse
with $p=k\rho$ are presented in subsection~\ref{prb-ss:mike}; in
particular, the critical exponents are tabulated for several values of
$k$ between zero and unity, and graphical evidence for self-similarity
in near critical collapse with $k=0.95$ is presented in
Fig.~\ref{prb-f:self-similar}.  Concluding the section on classical
results, we discuss the implications of critical point behavior for
cosmic censorship in subsection~\ref{prb-ss:cosmic-censorship}.  To fully
understand the physical significance of critical point behavior in
gravitational collapse requires the inclusion of quantum effects into
the picture.  Unfortunately, we do not have a complete theory of
quantum gravity, so the best we can do is address some model problems.
Several interesting results have been obtained in this context. Peleg
{\it et al.}\cite{Peleg_Y:1997} have investigated a 2-dimensional
dilaton theory of gravity in which they demonstrate the existence of
critical behavior and scaling at the classical level.  When one loop
quantum effects are included into the model, a mass gap is observed at
the threshold of black hole formation.  These results are reinforced
in similar work by Ayal and Piran,\cite{Ayal_S:1997} who have
considered a semi-classical model for spherically symmetric scalar
field collapse which is outlined in subsection~\ref{prb-ss:ayal}. General
arguments can be brought to bear on the four-dimensional problem when
the critical solution is continuously self-similar.  We present a brief
summary of a recent analysis by Brady and Ottewill~\cite{Brady_P:1998}
which suggests the existence of a quantum mass-gap at the
threshold of black hole formation for sufficiently stiff perfect
fluids.  Finally, we conclude with a brief discussion of
future directions.

%
%

\section{Classical results}\label{prb-s:classical}

\subsection{Massless scalar fields}\label{prb-ss:massless}

Since the initial discovery of critical phenomena, spherically
symmetric scalar field collapse has been one of the most studied
systems.  In contrast to perfect fluid collapse the critical solution
is not self-similar in the familiar sense, rather it has an infinite
train of self-similar echoes which accumulate at the singularity.
In precise terms, scale invariant quantities, such as $f$ in
Eq.~(\ref{prb-e:line-element}), satisfy
\begin{equation}
	f(\xi,\ln |t|) = f(\xi, \ln |t| - n \Delta /2)
\end{equation}
where $n$ is an integer and $\Delta \approx 3.4$ is the period
computed by Choptuik for the scalar field.  In words, scale invariant
quantities are periodic in $\ln |t|$ on surfaces of constant $\xi$.
The discrete self-similarity of the critical solution has implications
for the mass scaling law in Eq.~(\ref{prb-e:mscale}) --- there is periodic
fine structure superimposed on the power-law.  This was first noticed
by Gundlach,\cite{Gundlach_C:1997} and subsequently verified
numerically by Hod and Piran.\cite{Hod_S:1997}

A simple argument demonstrates this result.  The mass scaling law
arises because a single unstable mode governs the way that marginally
supercritical solutions run away from criticality. This mode can be
written as
\begin{equation}
	-(\eta-\eta_*) t^{-\lambda} f_1(\xi,\ln|t|)
\end{equation}
where $f_1(\xi,\ln|t|)$ inherits the symmetry of the background,
namely that it is periodic in $\ln |t|$.  Let $t_h$ satisfy
\begin{equation}
	f_* - (\eta-\eta_*) t_h^{-\lambda} f_1 = 0\; \label{prb-e:apparent1}
\end{equation}
for some fixed $\xi=\xi_h$.   Substitute $t_h=r_h/\xi_h$ into
Eq~(\ref{prb-e:apparent1}) and solve for $r_h$ to get 
\begin{equation}
	\ln r_h = \frac{1}{\lambda} \ln |\eta-\eta_*| + \Psi(\xi_h,\ln r_h)
	\; ,
\end{equation}
where $\Psi(\xi_h,\ln r_h)$ is periodic in $\ln r_h$ with period
$\Delta/2$.  If $\Psi\equiv 0$ we have a precise power-law relation,
however $\Psi$ is generally non-zero so an approximate solution for
$r_h$ is
\begin{equation}
	\ln r_h = \frac{1}{\lambda} \ln |\eta-\eta_*| + \Psi (\xi_h,
	 \lambda^{-1} \ln |\eta-\eta_*|) \; . \label{prb-e:periodic}
\end{equation}
Since $\Psi(\xi_h,\ln r_h)$ is periodic in $\ln r_h$,
Eq.~(\ref{prb-e:periodic}) implies that $r_h$, and equivalently the black
hole mass, has periodic fine structure with period $\lambda\Delta/2$
in $\ln|\eta-\eta_*|$ superimposed on the familiar power-law.

A natural extension of Choptuik's results which was considered by
Gundlach and Martin-Garcia,\cite{Gundlach_C:1996a} and also by Hod and
Piran,\cite{Hod_S:1997a} is scalar electro-dynamics coupled to
gravity.  The different scaling of the mass and charge in the
self-similar echoing regime suggests that the significance and
influence of the charge decreases during a near-critical evolution. As
$\eta-\eta_* \to 0$ the black-hole charge tends to zero faster than
its mass, i.e., $Q_{BH}/M_{BH} \to 0$ as $\eta \to \eta_*$ where
$Q_{BH}$ is the black hole charge. Gundlach and Martin-Garcia computed
the critical solution and predicted that the charge should scale as
\begin{equation}
	Q_{BH} \propto |\eta-\eta_*|^{0.883\pm0.007} \; .
\end{equation}
This scaling relation was confirmed by Hod and Piran in their
numerical simulations.

\subsection{Type I phase transitions}\label{prb-ss:massive}

By analogy with statistical mechanics, phase transitions in which
black-hole formation turns on at infinitesimal mass have been termed
{\em Type II} transitions.  Choptuik, Chmaj and
Bizon~\cite{Choptuik_M:1996} observed that black hole formation
sometimes turn on at a finite mass in the collapse of Yang-Mills field
coupled to gravity.  They called this a {\em Type I} transition since
the order parameter, the black hole mass, is not continuous at the
critical point.  The critical solution in this sector of the theory is
the Bartnik-McKinnon soliton, an unstable static solution of the
Einstein-Yang-Mills (EYM) equations.  The mass gap at the threshold of
black hole formation is approximately equal to the mass of the
Bartnik-McKinnon solution.  A fundamental difference between the EYM
system and the massless scalar field is the presence of a length
scale, the Yang-Mills charge.

\begin{figure}[h]
\centerline{
\psfig{file=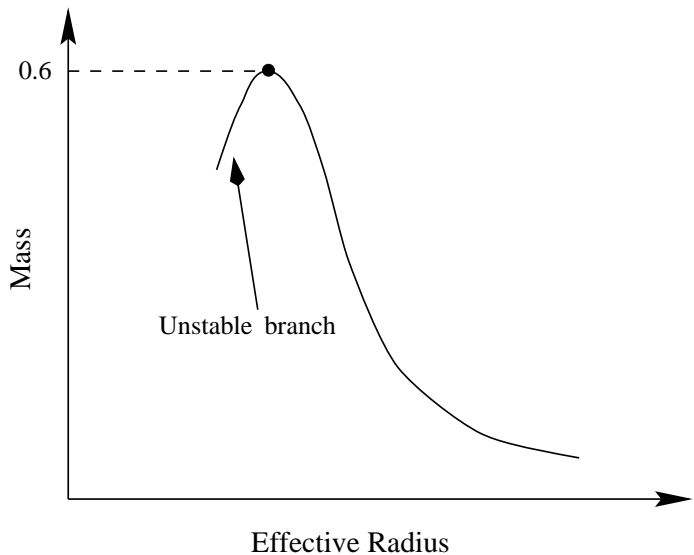,width=9cm,bbllx=120pt,bblly=247pt,%
bburx=491pt,bbury=544pt}
}
\caption{\label{prb-f:mass-radius}A schematic representation of the
mass-radius curve for oscillating soliton stars. (See Seidel and
Suen~\protect\cite{Seidel_E:1991} for a quantitatively correct version.)  The
maximum mass that one of these configurations can have is $\sim 0.6
\mu^{-1}$.  Solutions on the unstable branch of the mass radius curve
may act as critical solutions}
\end{figure}

It has been argued that Type II transitions in scalar field collapse
should be stable to the introduction of a mass.%
\cite{Gundlach_C:1995,Choptuik_M:1994,Hirschmann_E:1995c} Gundlach has
also presented a mathematical argument to this effect in his recent
review.\cite{Gundlach_C:1997d} Nevertheless, it was demonstrated by
Brady, Chambers and Gon\c{c}alves~\cite{Brady_P:1997a} that Type I
transitions can occur in massive scalar field collapse.  In contrast
to EYM collapse the mass gap at the threshold of black hole formation
is not universal; marginally supercritical black holes have masses
ranging from $\sim 0.3 \mu^{-1}$ to $\sim 0.6 \mu^{-1}$, where $\mu$
is the scalar field mass.  A plausible explanation for this is
provided by studying the critical solutions in Type I transitions;
they are oscillating soliton stars.  Seidel and
Suen\cite{Seidel_E:1991} have discussed these solutions in some
detail.  In particular, they demonstrated that there is a family of
solutions parameterized by the mass and the effective radius of the
star.  A schematic representation of the mass-radius curve is shown in
Fig.~\ref{prb-f:mass-radius}; note that the maximum mass of such a
star is $\sim 0.6 \mu^{-1}$, and solutions to the left of the maximum
are unstable.  The numerical evidence suggests that all of the
unstable solutions can act as critical point solutions, perhaps
explaining the range of values for the mass gap.

Why do some initial configurations lead to Type I transitions while
others lead to Type II?  A complete answer to this question is
unavailable at this time, however there is a simple criterion which
provides some guidance.  Let $\lambda$ denote the radial extent of the
initial shell of scalar field, {i.e.,}  its thickness.  Type I
transitions occur when the radial extent $\lambda$ of the initial
pulse is bigger than the Compton wavelength of the scalar field, that
is
\begin{equation}
	\lambda\mu \gg 1 \; .
\end{equation}
This observation, and the evidence in support of it that is presented
in Chambers' article,\cite{Chambers_C:thisvolume} validates local
arguments about the relevance of the mass term in Type II transitions.

It is likely that unstable, confined solutions play a role in
critical point behavior of other matter models.  Indeed, this is
expected in the astrophysical context where stars which exceed the
Chandrasekhar limit must either shed some material or collapse to form
a black hole.  The possible end-points of gravitational collapse in
this context are: (i) black hole formation, (ii) complete disruption
of the star, or (iii) a dead star with a mass less than $\sim 1.4
M_\odot$ (where $M_\odot$ is the mass of the sun).  In the most
general context, phase transitions between any pairwise combination of
these types can be expected, although all of the possibilities need
not occur.  Recent work by Chmaj and Bizon~\cite{Bizon_P:1998} lends
support to this viewpoint.  They have shown that stable skyrmion
solutions are a possible endpoint of gravitational collapse in Type I
transitions, {i.e.,}  phase transitions are observed between
stable stars and black holes.
 
\subsection{Brans-Dicke Theory}

Critical point behavior in Brans-Dicke theory has been studied by
Liebling and Choptuik.\cite{Liebling_S:1996} The critical solution
exhibits continuous self-similarity or discrete self-similarity
depending on the strength of the coupling of the Brans-Dicke field to
the matter (taken to be massless scalar field by Liebling and
Choptuik).  In a broader context, van~Putten has presented a
one-parameter family of static solutions to vacuum Brans-Dicke theory
which he suggests might be thought of as approximate black
holes.\cite{Vanputten_M:1996a} These solutions can be arbitrarily
close to the exterior Schwarzschild spacetime, but are globally
regular.  An important application of solutions like this could be to
provide approximate boundary conditions at the surface of a black
hole.  Choptuik, Hirschmann and Liebling~\cite{Choptuik_M:1997} have
argued that, in addition to the static solution behaving like a black
hole, the solution should also be stable.  Unfortunately, a linear
stability analysis indicates that van~Putten's solutions have an
unstable mode.  Using the code developed previously, Choptuik {\it et
al.} also studied the non-linear stability of these solutions.  They
evolved initial data corresponding to the static solution and small
additive perturbations.  Generically, the solution either collapses to
a black hole, or disperses depending on the sign of the perturbation.
An interesting point was raised by van Putten during the session on
critical phenomena at MG8.  He noticed, in animations of the perturbed
solutions, that the instability is evident only after the perturbation
reaches the center of symmetry and reflects back out toward larger
radii.  One may therefore wonder if the solutions might be stable when
absorbing boundary conditions would be applied at the origin; if so,
the solutions would become interesting as approximate black holes once
again.

Choptuik {\it et al.} have shown that van Putten's solutions have a
single unstable mode, and they either disperse or collapse when
perturbed.  This suggests that the solutions may be black-hole
threshold critical solutions.\cite{Koike_T:1995,Choptuik_M:1997} It is
intriguing to consider this possibility, since it would imply that
there is a basin of attraction in the space of initial data, for
massless scalar field collapse, in which van Putten's solutions
represent the critical solution.  Once again this begs the question:
what makes one solution a critical attractor, and another not?
Clearly, there is a gap in our understanding of critical phenomena.

\subsection{Perfect fluid collapse}\label{prb-ss:mike}

Despite the early study of critical phenomena in the collapse of
radiation fluid by Evans and Coleman,\cite{Evans_C:1994} no results
have been available on the evolution of perfect fluids with the
general equation of state $p=k\rho$, where $0<k\leq 1$ is a constant,
$\rho$ is the energy density of the fluid, and $p$ is the pressure.
Solutions expected to be at the threshold of black hole formation have
been computed by Maison~\cite{Maison_D:1996} and Koike {\it et
al.}\cite{Koike_T:1995} when $0<k\prbalt 0.899$.  In each case a single
unstable mode has been found, and a scaling exponent for the black
hole mass in slightly super-critical collapse has been predicted.

Brady and Cai~\cite{Brady_P:1998a} have developed spherically
symmetric code to study this problem.  The numerical scheme uses polar
slicing and a flux conservative version of the fluid equations of
motion which are differenced using a two-step Lax-Wendroff scheme.
The differencing scheme is second order accurate in both space and
time; this has been verified by performing a sequence of evolutions at
various levels of discretization.  The code was used to confirm the
results obtained by Evans and Coleman when $k=1/3$; at the threshold
of black hole formation self-similarity is observed over two orders of
magnitude in scale, and the critical exponent is computed to be
$\beta\approx 0.352$.

\begin{table}[h]
\caption{\label{t:exponents}
The scaling exponent for black hole mass for several values of $k$.
The values predicted using perturbation theory are in the second
column labelled $\beta_{\rm predicted}$.  The numerically observed
values and estimates of their errors are also presented}
\vskip 0.1in
\centerline{
\begin{tabular}{|c|c|c||c|c|c|}
\hline
$k$ & $\beta_{\rm predicted}$ & $\beta_{\rm observed}$ &$k$ &
$\beta_{\rm predicted}$ & $\beta_{\rm observed}$\\ \hline
0.1 & 0.1875 & 0.196 $\pm$ 0.009 &
0.6 & 0.5556 & 0.565 $\pm$ 0.011 \\
0.2 & 0.2614 & 0.265 $\pm$ 0.009 &
0.7 & 0.6392 & 0.653 $\pm$ 0.005 \\
0.3 & 0.3322 & 0.337 $\pm$ 0.002 &
0.8 & 0.7294 & 0.740 $\pm$ 0.003 \\
0.4 & 0.4035 & 0.408 $\pm$ 0.005 &
0.9 & --- & 0.814 $\pm$ 0.001 \\
0.5 & 0.4774 & 0.486 $\pm$ 0.004 &
0.95 & --- & 0.850 $\pm$ 0.005 \\ \hline
\end{tabular}
}
\end{table}

For equations of state with $0<k\prbalt 0.899$, the critical exponents
which have been computed from sequences of numerical evolutions agree
well with those obtained using perturbation methods.  Several values
are tabulated in Table~\ref{t:exponents}.  In near critical
evolutions, continuous self-similarity is observed over approximately
two orders of magnitude in scale before the matter either disperses or
collapses into a black hole.   

\begin{figure}[h]
\centerline{
\psfig{file=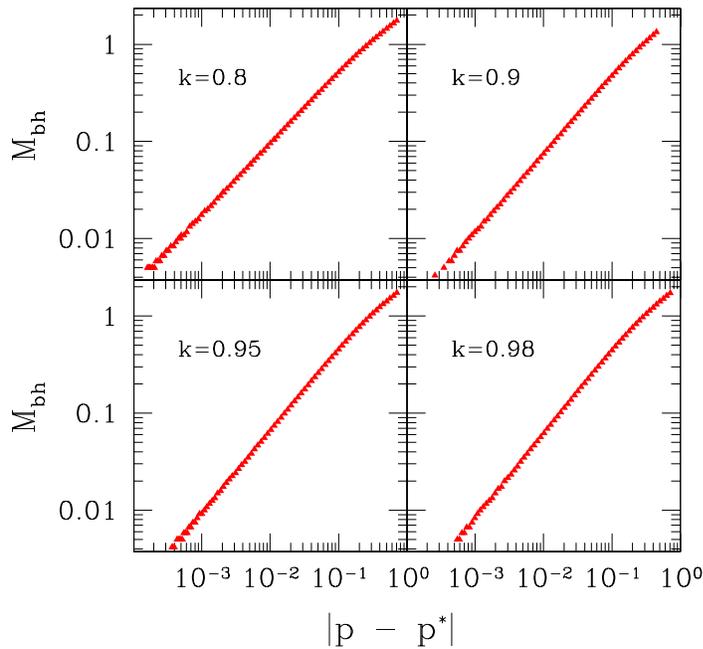,width=9cm,bbllx=48pt,bblly=184pt,%
bburx=592pt,bbury=698pt}
}
\caption{\label{prb-f:scaling}The scaling law for black hole mass in
marginally supercritical evolutions.  Results of a best fit to a power
law $M_{\rm BH} \propto |\eta-\eta_*|^\beta$ are presented for several
values of $k\geq 0.8$.  There is evidence for scaling over two orders
of magnitude in black hole mass;  the lower limit is set by the
resolution of runs which had $\Delta r = 0.001$.}
\end{figure}

The perturbative methods of Maison~\cite{Maison_D:1996} and Koike {\it
et al.}\cite{Koike_T:1995} cannot be applied when $k>0.899$ because
globally analytic, self-similar solutions representing perfect fluid
collapse fail to exist for such stiff equations of state.%
\cite{Ori_A:1990,Bogoyavlenskii_O:1978}
Nevertheless, Type II phase transitions are observed in numerical
simulations of collapse for these values of $k$.  The code allows
resolution of black holes down to $\sim 10^{-3} M$, where $M$ is the
ADM mass on the initial slice.  Figure~\ref{prb-f:scaling} shows the
scaling of black hole mass for several values of $k\geq 0.8$.

\begin{figure}[h]
\centerline{
\psfig{file=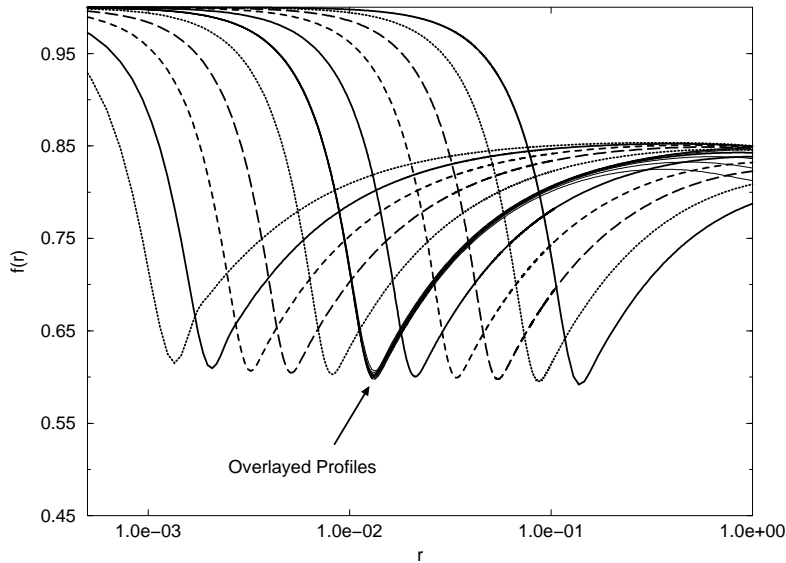,width=7cm,angle=-90,bbllx=94pt,%
bblly=186pt,bburx=544pt,bbury=606pt}
}
\caption{\label{prb-f:self-similar}Profiles of $f(r,t)$ when $k=0.95$ 
showing the self-similarity over approximately two orders of magnitude
in scale. The intermediate four profiles are overlayed when scaled in
accordance with the self-similar ansatz.}
\end{figure}

Near criticality, the solution develops a highly evacuated region
outside the collapsing matter, however there is evidence for
self-similarity over two orders of magnitude in scale in slightly
sub-critical solutions.  Several profiles of $f$ [see Eq.
(\ref{prb-e:line-element})] are shown in Fig~\ref{prb-f:self-similar}.  Some
of the profiles are overlayed when scaled according to the
self-similar ansatz; {\em continuous} self-similarity is apparent.
Since stiff fluid solutions can be recast as scalar field solutions,
one might expect that DSS should turn on as $k\rightarrow 1$.  No
evidence to this effect has been found in the numerical simulations.
The reason appears to be connected with a lack of analyticity at the
outer boundary of the collapsing material in perfect fluid collapse.

\subsection{Cosmic censorship and critical collapse}\label{prb-ss:cosmic-censorship}

The scaling law for black hole mass in Type II phase transitions
suggests that black holes of arbitrarily small mass can be formed in
gravitational collapse.  At the critical point, the central region of
collapsing material is self-similar; an infinite train of scale
invariant echoes accumulates within finite proper time in the echoing
solutions.  These two observations suggest that a singularity must
form at the center of critical solutions.  Is it naked?  Numerical
simulations by Hamade and Stewart~\cite{Hamade_R:1996b} provide direct
evidence that it is.  Using an evolution scheme based on double null
coordinates, they find a regular Cauchy horizon with almost vanishing
flux of scalar field across it.  Gundlach has analytically continued
the critical solution, constructed by his pseudo-spectral method, to
the Cauchy horizon which is regular.  Furthermore, self-similar,
perfect fluid solutions are known to have naked singularities at the
origin.\cite{Ori_A:1990} Thus, suitably chosen, regular initial data
can form a naked singularity in gravitational collapse; however, the
data are not generic in the sense that they belong to a sub-space of
codimension one in the space of all initial
data.\cite{Gundlach_C:1997d}

Weak cosmic censorship states that, in the evolution of generic
initial conditions, singularities are always hidden behind event
horizons.  Since critical solutions violate the generic condition,
they are not counter-examples to cosmic censorship.  Two further
assumptions are implicit in the above statement: First, gravitational
collapse is governed by the classical Einstein equations. Second,
matter is described by fundamental fields on spacetime.  Thus the
formation of naked singularities in the collapse of dust is not
usually considered a violation of cosmic censorship since dust only
provides an effective description of matter, and it can become
singular in flat spacetime.  Nevertheless, it is remarkable that naked
singularities are so easy to form in such matter.  Jhingan~{\it et
al.}\cite{Jhingan_S:1997} have examined dust collapse using the
Tolman-Bondi-Lema{\^i}tre metric, relating the formation of black
holes and naked, shell-focusing singularities in such a collapse to
the generic form of regular initial data. Such data are characterized
by the density and velocity profiles of the matter on some initial
time slice.  Given a generic initial density profile, they have shown
that there exists a corresponding velocity field which gives rise to a
strong curvature, naked singularity in the evolution. This establishes
that strong naked singularities arise for generic density profiles in
the spherically symmetric collapse of dust.

\section{Quantum effects in critical spacetimes}

It is well known, since the work of Hawking, that a black hole
decreases in size by emitting particles via quantum processes.  The
radiated particles have a thermal spectrum; for a static black hole
the temperature is inversely proportional to the mass of the black
hole.  Thus, the smaller the black hole, the more it radiates.  A
black hole of mass $M$ radiates all its mass in approximately
$10^{-27} (M/1\mathrm{g})^3$ seconds.  Black holes formed in marginally
super-critical collapse will therefore evaporate almost
instantaneously; quantum effects are important in a full description
of Type II phase transitions.

In the absence of a complete theory of quantum gravity, semi-classical
calculations can provide some information about the underlying quantum
evolution.  Since renormalization breaks conformal invariance, scale
invariant critical solutions surely are modified when curvatures
approach Planck scales.  Furthermore, the introduction of a
fundamental length scale, the Planck length, into the picture suggests
that a mass-gap might occur at the threshold of black hole formation
in a semi-classical treatment of critical point behavior.  These
speculations have been confirmed in several model problems.

\subsection{Two-dimensional dilaton models}\label{prb-ss:ayal}

Peleg {\it et al.}\cite{Peleg_Y:1997} have studied one-loop quantum
effects on the collapse of a massless scalar field in two-dimensional
(2D) dilaton gravity.  Reflecting boundary conditions are imposed at
some finite value of the dilaton $\phi=\phi_c$ in order to avoid the
strong coupling regime of the theory which has a timelike singularity
where $\exp(2\phi)\rightarrow\infty$.  The classical solutions exhibit
critical point behavior: in supercritical evolutions the black hole
mass $M_{\rm bh}$ scales as $M_{\rm bh} \propto |\eta - \eta_*|^{\beta}$
near to criticality.  In this 2D model, the critical exponent is
$\beta \simeq 0.53$. 

There is some freedom in constructing the effective action which
describes the semi-classical theory since one is free to add local
counter-terms to the Polyakov-Liouville term derived from the trace
anomaly in two dimensions.  Peleg {\it et al.} add a term which makes
the theory exactly soluble at the semi-classical
level.\cite{Bose_S:1996} The quantum coupling constant $\kappa= N\hbar
/12$ depends on the number of fields $N$.  As expected, quantum
effects are not relevant in the formation of sufficiently large black
holes.  The classical scaling relation holds over four orders of
magnitude when $\kappa = 0.001$, and over two orders of magnitude when
$\kappa = 0.01$.  The threshold of black hole formation is
characterized by a mass-gap in semi-classical evolutions, i.e.
$M_{\rm bh}$ approaches a {\it non-zero} lower limit as $\eta$ tends
to its critical value from above.  The mass-gap depends both on
$\kappa$, and on the initial data.  This is not surprising since the
semi-classical action is non-local.  Peleg has constructed analytic
arguments supporting these results (see his contribution to this
volume).

\subsection{Semi-classical models of spherical collapse}

The benefit of 2D models is that the quantum stress energy tensor can
be calculated exactly with very little effort.  Ayal and
Piran~\cite{Ayal_S:1997} have considered the four-dimensional,
semi-classical Einstein equations
\begin{equation}
	\mathbf{G} =  8\pi \left( \mathbf{T} + \left< \mathbf{T}
	\right> \right) \label{prb-e:semi-classical}
\end{equation}
where $\mathbf{G}$ is the Einstein tensor, $\mathbf{T}$ is the stress
energy of classical matter, and $\left<\mathbf{T}\right>$ is the
renormalized stress-energy tensor of quantum matter.  Since direct
computation of $\left<\mathbf{T}\right>$ is extremely difficult, if
not beyond current techniques, they use a model stress-energy tensor
constructed from a two-dimensional one computed on the radial two
sections~\footnote{By radial two section we mean the 2D spacetime
obtained by ignoring the angular dimensions in the spherical line
element.} of the four dimensional spacetime.  Tangential stresses are
ignored, and (in a slight abuse of notation) $4 \pi r^2
\left<\mathbf{T}\right> = [1+(\alpha/r^2)^2]^{-1}
\left< \mathbf{T}^{(2)} \right>$ where $\sqrt{\alpha}$ is a length scale of
order the Planck scale.  The prefactor is required so that the
stress-energy tensor is not singular at the origin in four dimensions.
Even though this term leads to small violations of covariant
conservation,  it is a useful model of semi-classical effects
including Hawking evaporation.

Solving the semi-classical equations (\ref{prb-e:semi-classical})
numerically for scalar field collapse, Ayal and Piran show that
quantum effects reduce the mass of a black hole compared to its
classical value.  When the length scale $\sqrt{\alpha}$ is increased
the black hole no longer forms.  As one might expect, the results are
qualitatively similar to those of Peleg {\it et al.}

\subsection{Four-dimensional model calculation}

An alternative approach exploits the symmetry of spacetimes with
continuous self-similarity to compute the general form of the
renormalized stress-energy tensor (RSET) for conformally coupled
fields.\cite{Brady_P:1998} Spherically, symmetric self-similar
spacetimes are conformally static, therefore one can apply the
transformation rule for the RSET, as derived by
Page,\cite{Page_D:1982} to show that
\begin{equation}
	\left< \mathbf{T} \right> = \hbar t^{-4} \left[ \mbox{ \bf functions
	of $\xi = r/t$ only} \right] \; . \label{prb-e:rset}
\end{equation} 
Here it is assumed that $t$ is the coordinate in
Eq.~(\ref{prb-e:line-element}) normalized so that $t=0$ at the
singularity in the critical solution.   

\begin{figure}[h]
\centerline{
\psfig{file=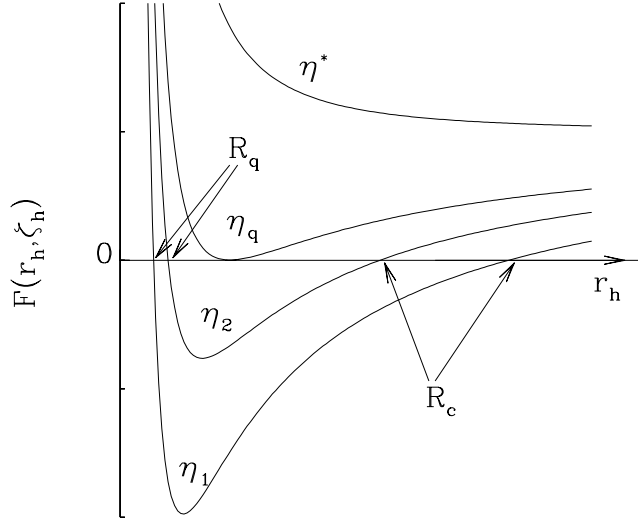,width=9cm,bbllx=18pt,bblly=244pt,bburx=592pt,bbury=718pt}
}
\caption{\label{prb-f:qroots}
The horizon location is determined by the roots of the function
$F(r_h,\xi_h)$ in Eq.~(\protect\ref{prb-e:qroots}).  We show here a
schematic representation for several values of $\eta$ which determine
deviations from classically critical initial data, and
$\lambda>2$.  For sufficiently large $\eta$ classical collapse
takes hold and a black hole forms at $R_c$.  The function has a
minimum, however, and another root $R_q$ exists.  As $\eta$ is tuned
to a critical value $\eta_q$ the two roots coincide.  When
$\eta<\eta_q$ no black hole forms.  }
\end{figure}

The critical solution in perfect fluid collapse is $f=f_*(\xi)$.  The
classical argument which relates the Lyapunov exponent to the scaling
relation for black hole mass can now be repeated including
perturbations which originate from the quantum stress-energy tensor in
Eq.~(\ref{prb-e:rset}).  The semi-classical equations imply that the
perturbations to the self-similar solution can be written as
\begin{equation}
	F(\xi,t) = f_* (\xi) - (\eta-\eta_*) f_1(\xi) t^{-\lambda} +
	\hbar f_q(\xi) t^{-2} \; , \label{prb-e:qroots}
\end{equation}
where the last term arises from quantum effects.
Clearly, when $\lambda < 2$ the quantum perturbations will dominate as
$t\rightarrow 0$.  Thus, we restrict attention to equations of state
with $k\prbagt 0.53$ for which $\lambda < 2$. (See Brady and
Ottewill~\cite{Brady_P:1998} for a discussion of $k<0.53$.)  Now,
substitute $t=r/\xi$ into Eq.~(\ref{prb-e:qroots}).  For fixed $\xi=\xi_h$
the solutions of
\begin{equation}
	F(\xi_h,r_h) = f_* (\xi_h) - (\eta-\eta_*) f_1(\xi_h)
	(r_h/\xi_h)^{-\lambda}  +
	\hbar f_q(\xi_h) (r_h/\xi_h)^{-2} =0 \; 
\end{equation}
determine the radii of apparent horizons in spacetime as a function of
$\eta$.  Assuming that quantum effects compete with classical
perturbations to reduce the mass of a black hole that forms in
gravitational collapse, $F(\xi_h,r_h)$ is plotted schematically in
Fig.~\ref{prb-f:qroots}. For $\eta$ sufficiently greater than $\eta_*$,
gravitational collapse and black hole formation is dominated by the
classical perturbations; the apparent horizon is at $R_c$.  Note that
the function has a minimum and a second root at $R_q<R_c$.  As $\eta$
decreases $R_c \rightarrow R_q$ until the roots coincide at some
$\eta_q$.  When $\eta<\eta_q$ no black hole forms; therefore, we infer
that quantum effects induce a mass gap at the threshold of black hole
formation in critical collapse of perfect fluids.  It is also apparent
that the classical scaling law for black-hole mass remains valid
for black holes with apparent horizons significantly above the Planck
length.

\section{Concluding remarks}

Critical point behavior in gravitational collapse is an exciting area
of research in general relativity, one which continues to bring new
insights into gravitational collapse and black hole formation.
Significant progress has been made to understand the phenomenology
observed in near critical evolutions.  Nevertheless, important open
questions remain.  For example, is cosmic censorship upheld in all
families of interpolating solutions for realistic matter fields?  Or,
when both Type I and Type II phase transitions occur, which properties
of the initial data determine the critical point behavior?  Numerical
studies of gravitational collapse can provide a great deal of insight
into the answers to these and other questions, however real progress
will be made at the interface between numerical and mathematical
approaches.

\section*{Acknowledgments}

PRB is grateful to Jolien Creighton, Sergio Gon\c{c}alves, Carsten
Gundlach, Eric Hirschmann, Scott Hughes, Adrian Ottewill, Leonard
Parker, and Joe Romano for useful conversations.  PRB extends his
warmest thanks to Chris Chambers for many stimulating discussions.
This work was supported in part by NSF Grant No. AST-9417371, the
Sherman Fairchild Foundation via a Caltech Prize Fellowship to PRB,
and a Summer Undergraduate Research Fellowship to MJC.


\end{document}